# Single/multi-site event discrimination of strip multi-electrode high purity germanium detector via pulse shape analysis method

**Yang Jingzhe** [a, b] **Zeng Zhi** [a, b] [*] **Dai Wenhan** [a, b] **Yang Mingxin** [a, b] **Tian Yang** [a, b] **Jiang Lin** [a, b] **Wen Jingjun** [a, b] **Xue Tao** [a, b] **Zeng Ming** [a, b] **Li YuLan** [a, b]

[a] *Department of Engineering Physics, Tsinghua University, Beijing 100084, China*
[b] *Key Laboratory of Particle and Radiation Imaging, Ministry of Education, Beijing 100084, China*
E-mail: zengzhi@tsinghua.edu.cn
* Zeng Zhi, E-mail: zengzhi@tsinghua.edu.cn

**Abstract**   In order to suppress the background in rare event detection experiments such as 0νββ, this paper developed a set of single/multi-site event pulse shape discrimination methods suitable for strip multi-electrode high-purity germanium detectors. In the simulation of $^{228}$Th, this method achieves 7.92 times suppression of SEP events at 2103 keV with a 57.43 % survival rate of DEP events at 1592 keV. The experimental study of $^{57}$Co and $^{137}$Cs sources was carried out by using the self-developed strip multi-electrode high-purity germanium detector prototype measurement system and compared with the simulation results. The results show that the discrimination effect of the PSD method on the experimental waveform is relatively consistent with that of the simulated waveform. The PSD method was used to identify the 0νββ background events of $^{76}$Ge. The survival rate of 0νββ events was 49.16 %, while the main background events $^{68}$Ge and $^{60}$Co, were 36.23 times and 31.45 times, respectively. The background suppression effects of $^{232}$Th and $^{238}$U were 4.79 times and 5.06 times, respectively. The results show that the strip multi-electrode high-purity germanium detector can be used for single/multi-site event discrimination and background suppression research. This method is expected to be applied in the measurement of 0νββ and other rare events in China Jinping Underground Laboratory in the future.

**Keywords:** Multi-electrode high-purity germanium; Pulse shape discrimination; Neutrinoless double beta decay; Background suppression

## 1 Introduction

Under the 'light Majorana neutrino' mechanism, the effective neutrino mass is inversely proportional to the 0νββ half-lives, and the absolute neutrino mass can be given by measuring the 0νββ half-lives[1]. High-purity germanium (HPGe) is both a target nucleus and a detector and is ideal for detecting 0νββ decay because of its high energy resolution, low background, and high detection efficiency. Several experiments have searched for 0νββ decay in $^{76}$Ge such as GERDA[2], MAJORANA COLLABORATION[3], and CDEX[4] experiments. The GERDA and the Majorana collaborations are now merged into the Legend collaboration and are proposing a 200 kg-scale 0νββ experiment (Legend-200) aiming at setting the 0νββ decay half-life limit of $^{76}$Ge at $10^{27}$ yr[5].

The 0νββ experiment is a typical ultra-low background physics experiment. In order to detect extremely rare signal events, the background needs to be suppressed to an ultra-low level. One of the key technologies to actively suppress the background is to analyze the pulse shape characteristics of the signal event and the background event, study the difference between the background and signal, and then discriminate the signal event to suppress the background event. This process is called pulse shape discrimination (PSD)[6]. In the 0νββ experiment based on HPGe, a large number of background events come from the partial energy deposition caused by Compton scattering of high-energy gamma rays[7]. The energy deposition is likely generated by multiple Compton scattering of γ-rays in HPGe crystals, so the background is mostly Multi-Site Event (MSE). On the other hand, the total energy deposition ($Q_{0\nu\beta\beta}$ = 2039 keV) region of two electrons from 0νββ decay is small, and the scale is about 1 mm, so the signal can be considered a Single-Site Event (SSE). If these two types of events are effectively distinguished, the experimental sensitivity of 0νββ can be improved[8].

The GERDA collaboration[9] and the MAJORANA COLLABORATION[10] developed waveform discrimination methods to discriminate single/multi-site events such as A/E. At the same time, other research groups have also developed PSD methods to identify single/multi-site events through pulse shape characteristics such as current waveform drop time[11] and rise time ratio[12]. However, these methods are all for point-contact-style HPGe detectors.

In recent years, multi-electrode HPGe detectors have been widely used in many fields because of their position sensitivity and good energy resolution. The AGATA experimental group used a segmented multi-electrode HPGe detector for track reconstruction research and measured the nuclear structure by waveform matching [13]. The COSI experiment used a strip multi-electrode HPGe detector for astronomical observations[14]. However, no relevant research has been conducted on single/multi-site event discrimination based on strip multi-electrode HPGe detectors and background suppression.

This study established a new single/multi-site event PSD method based on the strip multi-electrode high-purity germanium detector. We used the strip multi-electrode high-purity germanium detection system for background suppression research. Monte Carlo (MC) and pulse shape simulation (PSS) methods were used to study the effect of [228]Th source single/multi-site event discrimination. The experimental study of [57]Co and [137]Cs sources was carried out using the self-developed strip multi-electrode high-purity germanium detector prototype measurement system and compared with the simulation results. Finally, the suppression effect of the PSD method on the main background events in the 0νββ measurement of [76]Ge is evaluated.

## 2 Method

### 2.1 Multi-electrode high purity germanium experimental system and data acquisition

In this study, we used a multi-electrode HPGe detector developed by Tsinghua University[15]. We read the waveform signals of different electrodes by the evaporation of amorphous germanium on the surface of the HPGe crystal. Figure 1 shows the structure diagram of the crystal of the detector. The germanium crystal had a diameter of 30.0 mm and a height of 10.0 mm. Seven strips of p+ Al electrodes were located on the upper surface of the germanium crystal as signal readout electrodes with a width of 2.5 mm and a center distance of 0.5 mm. To reduce the influence of surface leakage current, we installed a 2-mm-wide protective ring outside the readout electrode. The bottom lithium electrode biased the detector, and the depletion voltage of the detector was +290 V. Considering the influence of surface leakage current, the recommended operating voltage was +300 V. This paper obtains the relevant physical parameters as the input of the subsequent simulation study through the established multi-electrode high-purity germanium detector experimental system.

As shown in **Figure 1**, seven readout electrodes were connected to a self-made nine-channel charge-sensitive preamplifier (CSP), and the output of each channel was connected to a reverse CR amplifier with zero-pole cancellation function. The CR output signal was then recorded and collected by a CAEN V1724 (14-bit) analog-to-digital converter (ADC) with a 100 MHz sampling rate. In the experiment, we collected 4000 samples (40 μs) for each waveform. The recorded original digital waveform is processed offline by PSA technology. The above two steps aim to ensure the basic quality of the original data. Firstly, the pulse shape is selected by checking the mean and linear slope of the baseline. Secondly, the stacking algorithm identifies and suppresses the superposition of multiple pulse shapes. Finally, all selected pulse shapes are baseline recovered and processed by the trapezoidal shaping algorithm to extract the energy information of the corresponding physical events for subsequent PSD analysis.

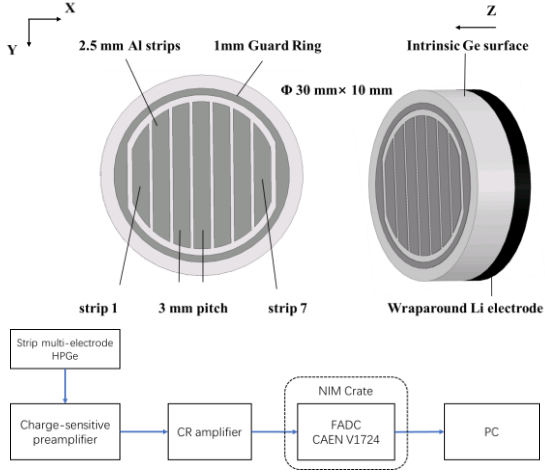

**Figure 1**(Left)The electrode distribution structure of the multi-electrode high-purity germanium crystal used in this paper. The inner diameter of the protective ring is 26 mm, and the seven strip-shaped induction electrodes are evenly distributed in a circle with a diameter of 24 mm. (Right)Data acquisition experimental system schematic diagram.

2.2 Single/multi-site event pulse shape discrimination method

In order to provide support for the study of the mechanism and waveform difference of single/multi-site events in strip multi-electrode high-purity germanium detectors and to simulate and evaluate the application effect of the PSD method, a waveform simulation method was established for the self-developed strip multi-electrode high purity germanium detector. The energy deposition of radiation particles in Ge crystals is calculated by the simulation toolkit SAGE[16] based on GEANT4[17]. The radiation particles' position and energy deposition information at each step level in the germanium crystal is obtained by Monte Carlo simulation. Taking the parameters shown in Section 2.1 as input, the electric field distribution and potential weight distribution in the detector are calculated by the open-source software SolidStateDector.jl[18]. Finally, according to the carrier drift trajectory in the detector combined with the Shockley-Ramo theory[19], the waveform corresponding to all the interaction points of the MC simulation output is added according to the energy deposition ratio, and the signal waveform generated by the electrode due to the charge collection process can be obtained. At the same time, as shown in the **Figure 2**, due to the structural characteristics of the multi-electrode high-purity germanium detector, the induced mirror signal waveform can also be calculated at other electrode positions. The mirror signal will generate a signal due to carrier motion during carrier drift and does not participate in the charge collection process, so the charge waveform of the mirror signal will eventually tend to zero. In summary, the signal waveforms induced on each electrode can be calculated separately.

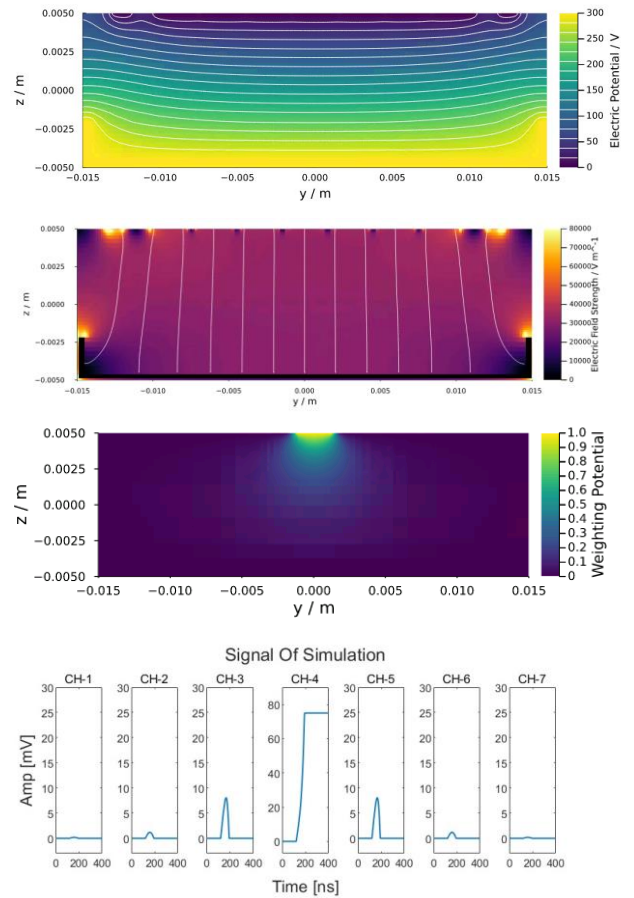

**Figure 2** The electric field and potential weight distribution in germanium crystal. (a) Potential distribution map; (b) Electric field distribution map; (c) The potential weight distribution of a single electrode on the upper surface; (d) Waveform readout schematics at different electrodes

For HPGe detectors, multi-site events usually interact at different positions far away from each other in the germanium crystal and the energy deposition area (about cm magnitude); single-site events usually have a smaller energy deposition area (about mm magnitude). Therefore, the signal waveforms collected by each readout electrode of the structure contain information about the event interaction, such as the location of the interaction and the total number of interactions. This study studied the waveform simulation and position resolution of the 2.1-section strip multi-electrode high-purity germanium detector. The position resolution of the multi-electrode high-purity germanium detector along the one-dimensional direction of electrode 1 to electrode 7 was calculated to be a sub-millimeter[20].

In this paper, a single/multi-site event pulse shape discrimination method for strip multi-electrode high purity germanium detector is developed. As shown in the **Figure 3**, for the strip multi-electrode high-purity germanium detector, due to its structural characteristics, when the multi-site event performs energy deposition at different radial positions, the generated charge signal will be collected by the corresponding readout electrode as the carrier drift process. Therefore, when performing single/multi-site event PSD, the number of induced electrodes for charge collection in the multi-electrode high-purity germanium detector is first judged. Suppose the final number of collected electrodes in the physical event is two or more multi-electrode collection (MEC) events. In that case, the event is regarded as a multi-site event to be eliminated.

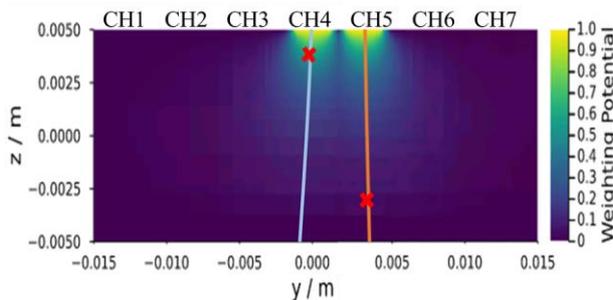

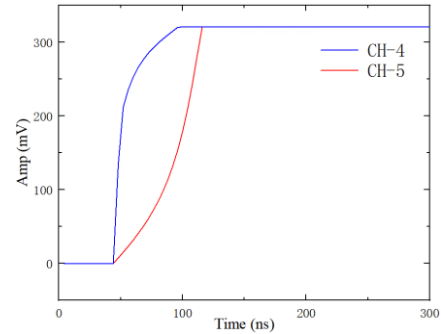

**Figure 3** Multi-electrode collection event diagram of multi-site events. (a)When Y = 0mm, energy deposition occurs at different positions in the X direction and is collected by different electrodes. (b) Multi-electrode collection event waveform diagram, two electrodes can sense significant charge waveforms and eventually collect charges.

In addition, as shown in the **Figure 4**, when the multi-site events interact at different axial positions, and the distance of the energy deposition point in the radial direction does not exceed the center distance of the strip electrode, the generated charge signal will be collected by the same readout electrode as the carrier drift process, becoming a single electrode (SEC) collection event. For the region below a single readout electrode, the weight potential distribution is axially symmetric with the readout electrode as the center. The weight potential distribution changes significantly near the readout electrode. In contrast, the potential weight gradient is smaller in the larger region away from the collector electrode. Currently, the potential weight distribution in this region is similar to that of single-electrode HPGe. According to the Shockley-Ramo law, the charge induced by the carrier drift to the electrode is proportional to the weight potential Wp at the position. Therefore, it is possible to consider using a pulse shape for single/multi-site pulse shape discrimination of events collected by a single electrode, such as the A/E method[9,10]. For single-site events, the peak value of the current waveform is approximately proportional to the total charge carried by the carrier. It is also proportional to the total energy deposited. For multi-site events, because the energy deposition is distributed at different interaction sites, the current waveform has multiple peaks, and the maximum peak is relatively small. In the case of the same energy deposition, the A/E value of the multi-site event waveform is smaller than that of the single-site event waveform. This feature can be used to identify single/multi-site events of single electrode collection to a certain extent. In addition, it is calculated that for the single site event acting on the electrode gap position on the surface of the germanium crystal, the region of the electrode gap in the geometric structure is compared with the region below the collector electrode.。

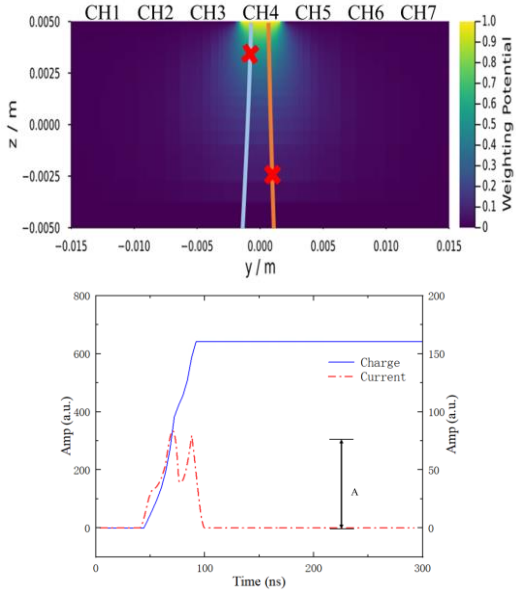

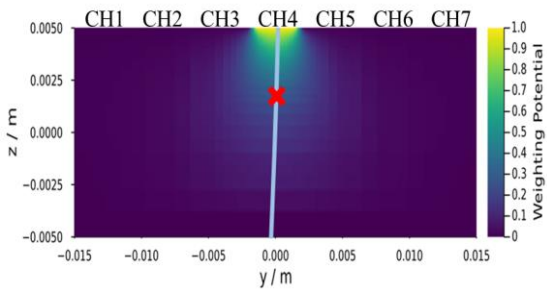

**Figure 4** Single-electrode collection diagram of multi-site events(a)When Y = 0mm, energy deposition occurs at different positions in the X direction and is collected by the same electrode. (b)Single-electrode collection event waveform diagram, a single electrode induces and collects charges, and the current waveform will have multiple peaks.

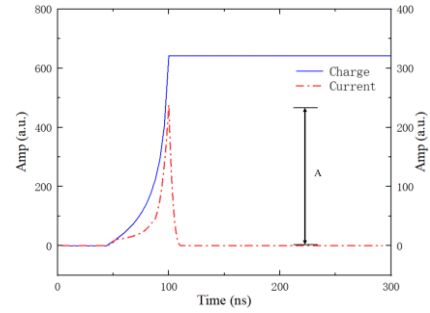

**Figure 5** Single electrode collection diagram of single site event (a)When Y = 0mm, a single energy deposition diagram occurs and is collected by the same electrode. (b)Single electrode collection event waveform diagram, a single electrode induces and collects charge, and the current waveform has only one peak.

## 3 Result and discussion

### 3.1 PSD method simulation study on single/multi-site event discrimination effect

$^{228}$Th is often used to scale the PSD method. Its daughter 208Tl can produce 2614keV γ-rays. The electron pair effect caused by the γ-rays in the germanium crystal is a typical SSE event (1592keV), and the corresponding single escape event (2103keV) is a typical MSE. The performance of the PSD method under the current system is often evaluated by analyzing the PSD method's discrimination rate of the double-escape peak (DEP)and single-escape peak (SEP)peaks. In this paper, the discrimination effect of the developed single/multi-site event PSD method is simulated and evaluated by simulating the $^{228}$Th source DEP and SEP event survival rate. The $^{228}$Th point source is set at 1cm above the center of the upper surface of the vacuum chamber of the 2.1 detection system, and the waveform simulation is performed on the simulated event. The A/E eigenvalues of DEP and SEP events are extracted by waveform analysis. The **Figure 6** shows that the A/E distribution corresponds to the 1592 keV DEP peak region event and the 2103 keV VSEP peak region event. It can be seen that there are differences in the distribution of A/E eigenvalues between the two events. Most DEP peak area events are distributed in Gaussian shape characteristic peaks with higher A/E values, and there are a small number of

counts at low A/E values. For SEP peak events, in addition to a Gaussian-shaped characteristic peak, the distribution also has obvious counts at low A/E values due to the contribution of multi-site events in the energy region. At the same time, according to the difference in A/E distribution, for the multi-electrode high-purity germanium detector with the current structure, the A/E discrimination threshold of the single electrode collection event is set to 0.10, and the events less than the discrimination threshold are eliminated to realize the suppression of multi-site events in the background.

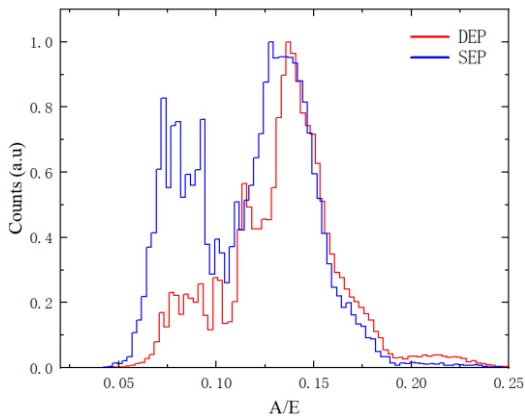

**Figure 6** The simulated $^{228}$Th sources A/E value distribution map. The red was the DEP event A/E distribution, and the blue was the SEP event A/E distribution.

The developed PSD method discriminated single/multi-site events, and the survival rates of 1592keV DEP events and 2103keV SEP events before and after cut were calculated, respectively. As shown in the **Table 1**, it can be seen that the multi-electrode high-purity germanium measurement system of this structure can significantly suppress SEP multi-site events under ideal conditions, and the net count is suppressed by 7.92 times after cut. At the same time, the survival rate of the DEP single event was 58.43 %. In addition, as shown in the **Figure 7**, blue is the energy spectrum after the single electrode collection cut of the original energy spectrum, and red is the energy spectrum after the further A/Ecut of the single electrode collection event. Due to the structural characteristics of the multi-electrode high-purity germanium detector, the background can be suppressed to a certain extent by collecting the number of electrodes without analyzing the difference in waveform shape.

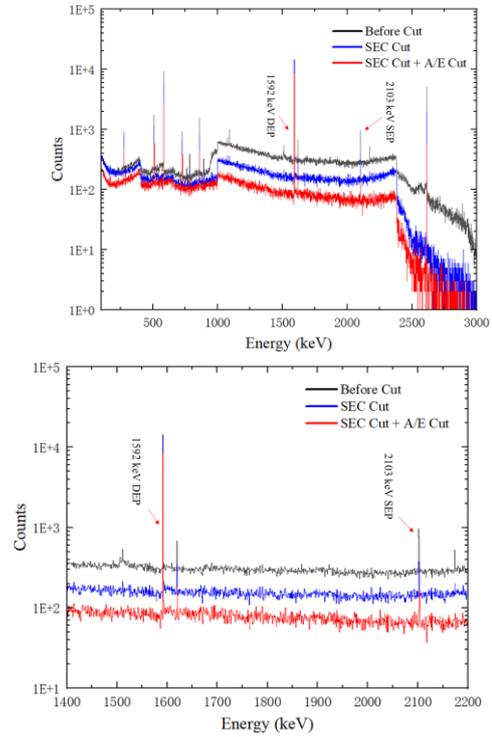

**Figure 7** Energy spectrum before and after $^{228}$Th source discrimination. The blue is the energy spectrum collected by a single electrode after Cut, and the red is the energy spectrum collected by a single electrode after A/E cut (a)the simulated energy spectrum of $^{228}$Th before and after cut (0-3000keV) (b)the simulated energy spectrum of $^{228}$Th before and after cut (1400-2200keV).

**Table 1** $^{228}$Th source before cut and after cut the peak net count acceptance rate results.

|  | DEP (1592 keV) | SEP (2103 keV) |
|---|---|---|
| SEC cut Acceptance | 96.92% | 33.95% |
| SEC cut + A/E cut Acceptance | 58.43% | 12.62% |

3.2 Experimental study of the PSD method

Through MC simulation, it is found that the 122 keV all-energy peak event of the $^{57}$Co source has a high probability of depositing all energy directly with germanium crystal due to its low energy, so single-site events dominate the all-energy peak event. The 661 keV all-energy peak event of the $^{137}$Cs source has a high probability of several Compton scatterings with the germanium crystal and, finally, the photoelectric absorption deposition of all energy, so the multi-site event of the all-energy peak event accounts for a large proportion. Therefore, $^{57}$Co and $^{137}$Cs point sources can be measured experimentally, and the PSD method can be used to conduct single/multi-site event discrimination research on the collected waveform data. In this paper, through the prototype measurement system shown in Section 2.1, the $^{57}$Co and $^{137}$Cs point sources are placed 1cm above the center of the upper surface of the vacuum chamber for measurement. Finally, 6w group $^{57}$Co waveforms and 16w group $^{137}$Cs waveforms are collected. As shown in the **Figure 8**, the survival rate of the 122 keV all-round peak of $^{57}$Co is $(82.52 \pm 0.04)\%$, while the 661 keV all-round peak of $^{137}$Cs is suppressed by about five times. In addition, this paper also simulates the two types of point sources, respectively, and the simulation discrimination effect of the two types of point sources by the PSD method is shown in the **Table 2**. Through comparison, it can be seen that the discrimination effect of the PSD method on experimental waveforms is relatively consistent with that of simulated waveforms.

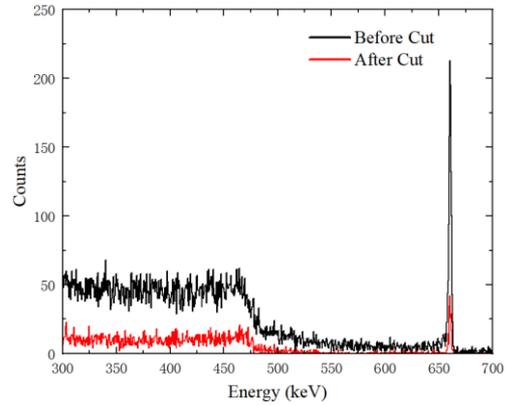

**Figure 8** Experimental energy spectra before and after $^{57}$Co source and $^{137}$Cs source discrimination. Black is the original energy spectrum, and the energy spectrum after the red PSD method (a)discriminates the $^{57}$Co source experimental energy spectrum before and after the discrimination (b)discriminates the $^{137}$Cs experimental energy spectrum before and after the discrimination.

**Table 2** The net count acceptance rate results of energy peaks before and after discrimination of $^{57}$Co source and $^{137}$Cs source.

|  | $^{57}$Co（122 keV） | $^{137}$Cs（661 keV） |
| --- | --- | --- |
| Experiment Acceptance | $(82.52\pm0.04)\%$ | $(20.44\pm0.45)\%$ |
| Simulation Acceptance | 82.49% | 20.17% |

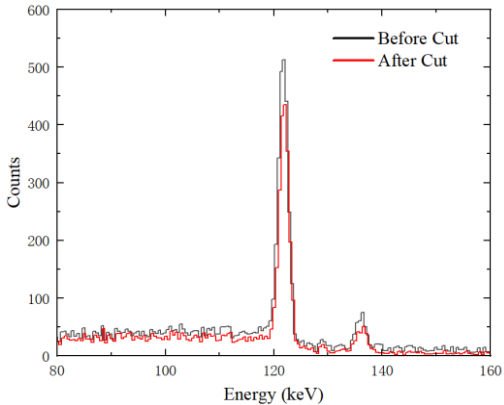

### 3.3 Evaluation of discrimination efficiency of 0νββ background events

This paper uses the PSD method of strip multi-electrode high purity germanium detector to suppress the 0νβ background event of $^{76}$Ge. In order to simulate the 0νββ event, this paper uses SAGE for MC simulation. In the simulation, 0νββ events are set up to perform uniform volume sampling in HPGe crystals and generate emission electrons while cut events that retain all energy deposition in HPGe. In addition, the main background source terms in the 0νββ measurement of $^{76}$Ge are simulated: $^{68}$Ge, $^{60}$Co, $^{232}$Th, $^{238}$U, and the event information in the energy range 1900keV-2100keV is counted.

The PSD method is used to identify the single/multi-site events of the 0νββ signal event and the main background event, and the results are shown in the **Table 3**. Due to the relatively small deposition area of the total energy ($Q_{0\nu\beta\beta}$ = 2039 keV) of the two electrons in the 0νββ event, the vast majority of them are still retained as single-site events after cut, and the survival rate is 49.16 %. The background events are suppressed after cut due to the relatively long distance of interaction energy deposition. Among them, the background suppression effects of $^{68}$Ge and $^{60}$Co sources are more significant, 36.23 times and 31.45 times, respectively. In comparison, the background suppression effects of native $^{232}$Th and $^{238}$U sources are 4.79 times and 5.06 times, respectively. This is because the high-energy γ events produced by primary nuclides have a certain probability of depositing energy only through one Compton scattering in this energy interval and becoming the background source. The event of cosmogenic nuclide production is a typical multi-site event in which electrons and γ-rays are released through decay to deposit energy in the detector.

**Table 3** The acceptance rate results before and after the cut of 0νββ signal events and main background events.

| | 0νββ | $^{68}$Ge | $^{60}$Co | $^{232}$Th | $^{238}$U |
|---|---|---|---|---|---|
| Acceptance | 49.16% | 2.76% | 3.18% | 19.77% | 20.89% |

## 4 Conclusion

In this paper, a single/multi-site event PSD method suitable for strip multi-electrode high purity germanium detector is developed, and the strip multi-electrode high purity germanium detector is used for background suppression research. In the simulation of $^{228}$Th, this method achieves 7.92 times suppression of SEP events at 2103 keV with a 57.43 % survival rate of DEP events at 1592 keV. The $^{57}$Co and $^{137}$Cs sources were measured by the self-developed strip multi-electrode high-purity germanium detector measurement system. The results show that the discrimination effect of the PSD method on the experimental waveform is relatively consistent with that of the simulated waveform. Finally, the PSD method was used to discriminate 0νββ background events of $^{76}$Ge. The results showed that most of the 0νββ events were still retained as single-site events after the cut, and the survival rate was 49.16 %. The main background events were suppressed after the cut, and the background suppression effects of $^{68}$Ge and $^{60}$Co were 36.23 times and 31.45 times, respectively. The background suppression effects of the method on the primary nuclides $^{232}$Th and $^{238}$U were 4.79 times and 5.06 times, respectively.

It can be seen from the research that the multi-electrode high-purity germanium detector can perform single/multi-site event discrimination, which can provide a new means for background suppression of rare event measurements such as 0νββ in the future. In addition, it is found that the strip multi-electrode high-purity germanium detector can suppress the background to a certain extent by collecting the number of electrodes without analyzing the difference in waveform shape due to its structural characteristics. With the development of the development technology of multi-electrode HPGe detector, the PSD background suppression method developed in this paper is expected to play a more important role in the measurement of rare events in China Jinping Underground Laboratory in the future.

**Acknowledgments**
This work was supported by the National Natural Science Foundation of China (Grants No. U1865205).